\documentclass[sigconf]{acmart}

\usepackage{booktabs}
\usepackage{array}
\usepackage{tabularx}

\usepackage{xspace}
\newcommand{\tool}{\textsc{ImaginalExpoBot}\xspace}

\AtBeginDocument{%
  \providecommand\BibTeX{{%
    \normalfont B\kern-0.5em{\scshape i\kern-0.25em b}\kern-0.8em\TeX}}}

\setcopyright{acmcopyright}
\copyrightyear{2026}
\acmYear{2026}
\setcopyright{cc}
\setcctype{by}
\acmConference[IH '26]{Interactive Health Conference}{July 05--08, 2026}{Porto, Portugal}
\acmBooktitle{Interactive Health Conference (IH '26), July 05--08, 2026, Porto, Portugal}
\acmDOI{10.1145/3786579.3804922}
\acmISBN{979-8-4007-2422-0/2026/07}

\begin{document}

\title{Explore LLM-enabled Tools to Facilitate Imaginal Exposure Exercises for Social Anxiety}


\author{Yimeng Wang}  
\affiliation{
    \department{Computer Science}
    \institution{William \& Mary}
    \city{Williamsburg}
    \state{VA}
    \country{USA}}
\email{ywang139@wm.edu}
\orcid{0009-0005-0699-4581} 

\author{Yinzhou Wang}  
\affiliation{
    \department{Computer Science}
    \institution{William \& Mary}
    \city{Williamsburg}
    \state{VA}
    \country{USA}}
\email{ywang143@wm.edu}
\orcid{0009-0009-6355-2551} 

\author{Alicia Hong}  
\affiliation{
    \department{Dept of Health Administration Policy and Informatics}
    \institution{George Mason University}
    \city{Fairfax}
    \state{VA}
    \country{USA}}
\email{yhong22@gmu.edu}
\orcid{0000-0002-1481-6495} 

\author{Yixuan Zhang}  
\affiliation{
    \department{Computer Science}
    \institution{William \& Mary}
    \city{Williamsburg}
    \state{VA}
    \country{USA}} 
\email{yzhang104@wm.edu}
\orcid{0000-0002-7412-4669}

\renewcommand{\shortauthors}{Wang et al.}

\begin{abstract} 
Social anxiety (SA) is a prevalent mental health challenge that significantly impacts daily social interactions. Imaginal Exposure (IE), a Cognitive Behavioral Therapy (CBT) technique involving imagined anxiety-provoking scenarios, is effective but underutilized, in part because traditional IE homework requires clients to construct and sustain clinically relevant fear narratives. In this work, we explore the feasibility of an LLM-enabled tool that supports IE by generating vivid, personalized exposure scripts. We first co-designed \tool with mental health professionals, followed by a formative evaluation with five therapists and a user study involving 19 individuals experiencing SA symptoms. Our findings show that LLM-enabled support can facilitate preparation for anxiety-inducing situations while enabling immediate, user-specific adaptation, with scenarios remaining within a therapeutically beneficial ``window of tolerance''. Our participants and MHPs also identified limitations in continuity and customization, pointing to the need for deeper adaptivity in future designs. These findings offer preliminary design insights for integrating LLMs into structured therapeutic practices in accessible, scalable ways.
\end{abstract} 






\maketitle

\section{Introduction} 

Social anxiety is an increasingly prevalent mental health issue, exacerbated by the social disruptions due to the COVID-19 pandemic~\cite{thompson2021changes, liang2021impact, ambusaidi2022prevalence}. Individuals with social anxiety often face heightened challenges in engaging with others, fearing judgment, embarrassment, or rejection in everyday interactions~\cite{morrison2013social}. Cognitive Behavioral Therapy (CBT)~\cite{kaczkurkin2015cognitive} is widely recognized as an effective approach for coping with social anxiety. Within CBT, imaginal exposure (IE) has emerged as a valuable technique for helping individuals confront and process anxiety-provoking situations~\cite{ledley2009comprehensive}. IE works by guiding individuals to vividly and repeatedly imagine a feared scenario in detail---including sensory elements (sights, sounds, smells), emotional responses, and cognitive appraisals---while remaining within a ``window of tolerance,'' a manageable level of discomfort that supports therapeutic engagement~\cite{corrigan2011autonomic}. Through repeated mental rehearsal, individuals gradually habituate to the anxiety response, reducing avoidance behaviors over time~\cite{heimberg1985treatment}. 

IE is often practiced both in sessions and as \textit{``homework''}, typically by writing a detailed script describing one's fears and then listening to a recording of it daily~\cite{ledley2009comprehensive}. However, these traditional methods can be labor-intensive, relying heavily on the individual’s ability to actively generate, elaborate, and sustain detailed fear narratives during exposure~\cite{schwob2023brief}. Some individuals struggle to produce vivid scripts, while others find it difficult to maintain consistent engagement without immediate support. These barriers limit the accessibility and effectiveness of IE homework~\cite{thompson2021changes}. Recent advancements in Large Language Models (LLMs) offer a promising opportunity to address the challenges. LLMs can generate vivid, context-rich descriptions and adapt dynamically to user input, capabilities that align with IE’s emphasis on detailed mental imagery. While LLMs have shown promise in other CBT modalities~\cite{Jiang2024}, such as cognitive distortion detection~\cite{chen2023empowering} and cognitive restructuring~\cite{sharma2023cognitivereframingnegativethoughts}, their potential for supporting IE remains unexplored. In this work, we explore the following research questions: 
\begin{itemize}
    \item[\textbf{RQ1.}] How can LLM-enabled tools be designed to support IE exercises for individuals with social anxiety? and
    \item[\textbf{RQ2.}] What are the perceived benefits and limitations of LLM-enabled IE tools from the perspectives of both users and mental health professionals (MHPs)?
\end{itemize}

To address these questions, we co-designed an LLM-enabled chatbot, ImaginalExpoBot, in collaboration with MHPs. We then conducted a formative evaluation with five therapists and a user study with 19 participants experiencing social anxiety symptoms. Our results show a strong preference for common social scenarios (e.g., parties, job interviews) when practicing imaginal exposure using \tool, and suggest that participants perceived \tool as helpful in preparing for real-life anxiety-provoking situations by providing vivid, multi-sensory scripts that mirror established IE practices. Both users and MHPs praised its ability to create strikingly realistic narratives and maintain anxiety within a ``window of tolerance,'' ensuring discomfort remains engaging rather than overwhelming. These features, coupled with real-time adaptivity, suggest promise for reducing avoidance behaviors and enhancing compliance with IE homework. Participants and MHPs also noted that \tool sometimes failed to carry context forward, leading users to repeat surface-level concerns rather than uncover deeper anxieties. They also highlighted the one-size-fits-all design, which offered limited customization and felt rigid rather than immersive or adaptive. 
This work \textbf{contributes}: (1) an exploration of how LLM-enabled tools can be designed to support IE exercises, and (2) an evaluation examining the potential and limitations of such tools, with implications for designing health-centered, interactive LLM systems.

\section{Background \& Related Work}
\subsection{Social Anxiety and Imaginal Exposure}

Social anxiety has become an increasingly prominent issue, especially in the aftermath of the COVID-19 pandemic, as prolonged isolation and limited social interaction have heightened anxiety levels in social contexts. CBT~\cite{beck2011cognitive} has been effective in addressing social anxiety, and IE offers a controlled approach by guiding individuals to imagine anxiety-provoking scenarios while remaining within their ``window of tolerance''~\cite{corrigan2011autonomic}. IE is typically implemented both during therapy sessions and as homework, where clients write about imagined scenarios or repeatedly listen to a recording of their scripts to facilitate emotional processing and habituation~\cite{heimberg1985treatment}. 

However, these traditional approaches can be labor-intensive and cognitively demanding. Some individuals struggle to produce vivid scripts, leading to incomplete or less effective practice, while others find it difficult to maintain consistent engagement without immediate support or prompts~\cite{nuts_and_bolts_of_imaginal_exposure}. Therapists, in turn, face challenges in monitoring homework exercises. These limitations motivate our exploration of LLMs to enhance IE’s adaptability and accessibility.

\subsection{LLM-enabled Cognitive-Behavioral Therapy}

LLMs have gained significant traction in digital psychotherapy due to their accessibility and constant availability~\cite{obradovich2024opportunities, Wang2025}. Recent research has explored LLMs for detecting cognitive distortions (i.e., maladaptive beliefs instead of objective facts)~\cite{chen2023empowering}, facilitating cognitive restructuring (i.e., reframing cognitive distortion in to balanced thoughts)~\cite{sharma2023cognitivereframingnegativethoughts}, and simulating patients with cognitive models~\cite{wang2024patient}. 
Recent work has also explored integrating LLMs into structured therapies through prompt engineering~\cite{filienko2024toward} and script-strategy aligned generation~\cite{sun2024script}.
While existing LLM-based exposure-related technologies, such as multimodal VR design systems, demonstrate promising interfaces, they have overlooked the challenge of generating personalized and context-sensitive imaginal exposure scripts for everyday users~\cite{Yuta2025}. IE exercises require precisely this level of personalization and context sensitivity: what matters most is not which scenario is chosen, but how the script captures each individual's unique experiences, sensory associations, and anxiety triggers within that scenario, while keeping the experience within a therapeutically tolerable range. To the best of our knowledge, this is the first work to explore LLM-enabled tools specifically designed for IE exercises, bridging an important gap in the intersection of interactive health technologies.

\section{Methods}
\label{sec:methods}

Below, we describe the co-design of \tool, our evaluation studies, and data analysis approach. This research was approved by the Institutional Review Board (IRB) at our institution.

\subsection{\tool Design and Co-Design Process}
\label{sec:IE_design}

\begin{figure*}[h]
    \centering
    \includegraphics[width=0.9\textwidth]{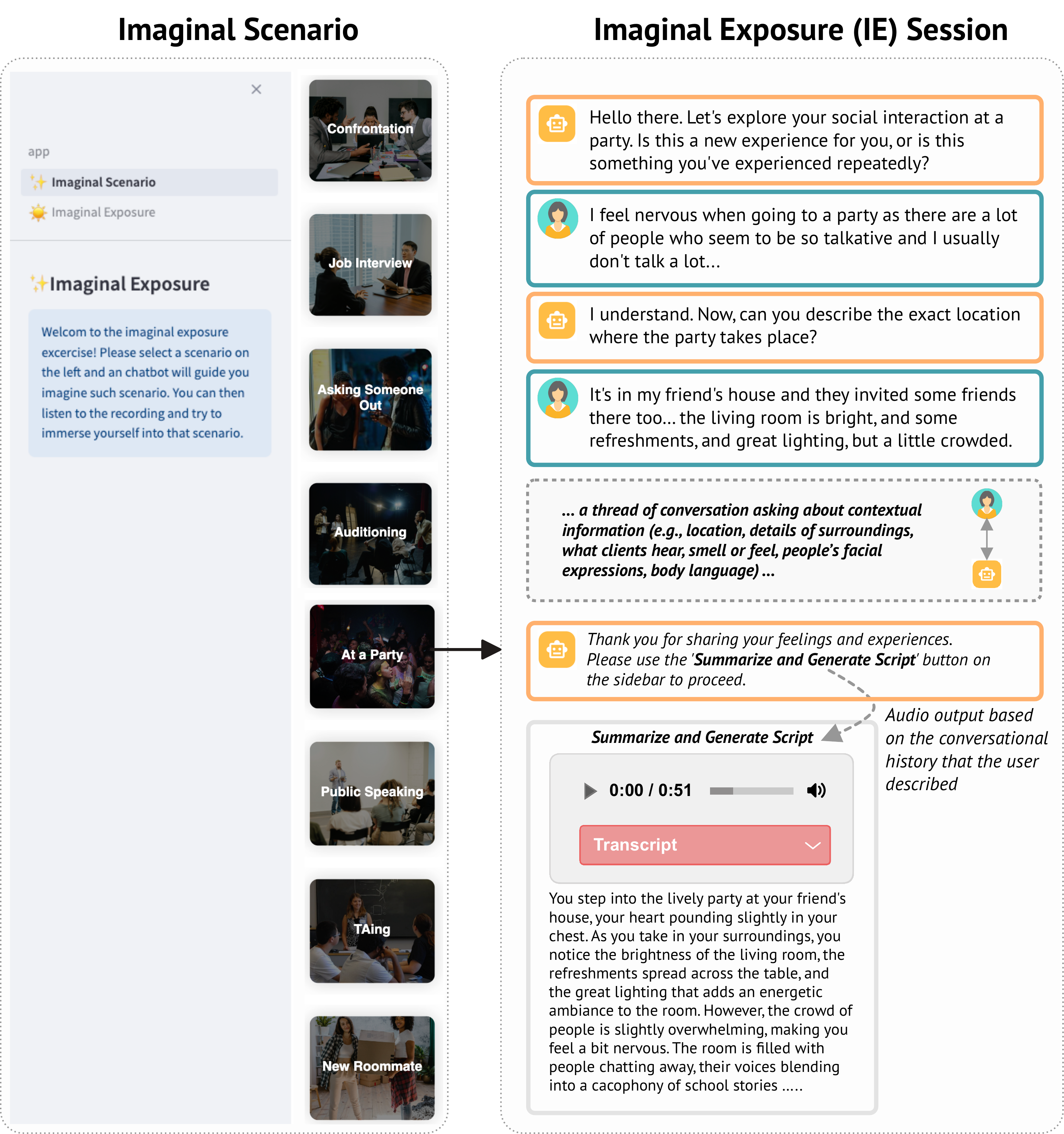}
    \caption{An illustration of the logic and conversational flow of the Imaginal Exposure. Users first choose one of the eight imaginal scenarios (left) and then start with the IE practice (right). }
    \label{fig:IE_flow} 
\end{figure*} 

We designed \tool through iterative collaboration with five mental health professionals (MHPs), each with active clinical caseloads and an average of 15 years of counseling experience (range: 5–35 years). During the co-design phase, we conducted structured consultations in which MHPs reviewed common IE prompts used in their clinical practice, identified key elements of effective IE scripts (e.g., multi-sensory detail, emotional cues), and provided feedback on interaction flows. Their input shaped both the scenario selection and the prompting strategy.

As shown in \autoref{fig:IE_flow}, \tool guides users through a two-phase process. In \textbf{Phase 1 (Scenario Selection)}, users choose from eight predefined social anxiety scenarios: confrontation, job interviews, attending a party, public speaking, asking someone out, mentoring students as a TA, auditioning, and meeting a new roommate. These scenarios were selected based on the social anxiety literature~\cite{hur2020social, leary1997social} and MHP input regarding common triggers seen in their clinical practice. In \textbf{Phase 2 (Guided Script Development)}, an LLM agent engages the user in a structured conversation to build a personalized IE script. The agent asks a sequence of targeted questions, eliciting information about: (1) past experiences with the scenario, (2) the setting and environment, (3) sensory details such as what the user sees, hears, or smells, (4) interactions with other people in the scenario, and (5) the user's feelings and thoughts throughout. This prompting sequence was directly informed by CBT guidelines that emphasize multi-sensory detail in IE scripts~\cite{ledley2009comprehensive} and refined based on the MHPs' recommendations (e.g., emphasize everyday, grounded details rather than catastrophic outcomes). The conversation typically involves 10–15 exchanges, with users providing varying levels of detail. Once sufficient information is gathered, the LLM synthesizes all inputs into a cohesive imaginal script written in the second person and present tense (e.g., ``You step into the crowded room...''). The script is then converted to audio using a text-to-speech model, allowing users to repeatedly listen to the recording (a core IE homework practice~\cite{ledley2009comprehensive}).

\subsection{Evaluation Study}
\label{sec:interview}

\subsubsection{Formative Evaluation with MHPs.}
After prototyping the system, we conducted one-hour formative evaluation sessions via Zoom with the same five MHPs. Each session began with a tool demonstration by the lead author. MHPs then independently explored the platform using the think-aloud method, verbalizing their thoughts and reactions as they navigated the interface~\cite{van1994think}.

\subsubsection{User Study with 19 Participants. } 

We leveraged a participant pool from the research team’s previous projects on social anxiety. Email invitations were sent to approximately 150 individuals who had previously reported social anxiety symptoms or related concerns. The email included a detailed description of the study’s purpose, procedures, a consent form, and a screening survey (see Survey Details in the Appendix). Several safeguards were implemented to ensure responsible use. The system was explicitly framed as a supplement to, not a substitute for, clinical therapy. During onboarding, participants were informed of available mental health resources. Throughout the study, licensed therapists were available for consultation if participants experienced heightened anxiety. Additionally, participants with ``severe'' scores on validated screening measures (SPIN~\cite{connor2000psychometric}, PHQ-9~\cite{kroenke2001phq}, and GAD-7~\cite{spitzer2006brief}) were excluded from the study to ensure participant safety. 19 participants (9 men, 10 women, mean age = 20, $SD =$ 1.10, range = 18–23) completed the user study and exit survey. The sample was predominantly White ($n =$ 9), followed by Asian ($n =$ 7), Multiracial ($n =$ 2), and one participant who did not report their race. Three participants identified as Latino/Hispanic. Education levels included high school diploma or equivalent ($n =$ 9), some college/associate degree ($n =$ 11), and bachelor’s degree ($n =$ 2). Participants reported mild-to-moderate symptoms with average scores of 20.33 (SPIN, $SD =$ 14.23), 5.57 (PHQ-9, $SD =$ 4.36), and 5.19 (GAD-7, $SD =$ 3.36).

\textbf{User evaluation procedures.} First, participants attended an onboarding session via Zoom, during which a researcher introduced the study's goals and procedures, clarified that \tool was not a substitute for clinical therapy, and emphasized the availability of mental health resources if distress arose. A demonstration video familiarized participants with the features and functionality of \tool. Participants were encouraged to ask questions during the session to address any concerns and ensure they felt comfortable with the study requirements. After onboarding, participants independently used \tool at their own pace and convenience over a two-week period in the wild. Users were allowed to explore \tool’s features and functionalities in real-life scenarios, providing opportunities to integrate the tool into their daily lives and test its practical applicability. Upon completing the interaction period, participants completed an exit survey regarding their feedback on their experiences, perceived changes in anxiety, and suggestions for improvements. Each participant received a \$40 gift card.

\subsection{Data Analysis} 
We collected data from surveys, MHP interviews, and system log files. Descriptive statistics were calculated to summarize key metrics. For qualitative data analysis, we used the General Inductive Approach~\cite{thomas2006general} to guide the thematic analysis of open-ended survey responses. The first author closely reviewed the free-text responses to gain an initial understanding of emerging concepts. Low-level codes were created to label these concepts, which were then grouped into related clusters to develop high-level themes. Throughout the analysis, all authors engaged in regular discussions to refine and validate the emerging themes.

\section{Results}

Participants completed a total of 88 IE exercises across the study. The most frequently chosen scenarios were ``Party'' ($n =$ 18, 20\%), ``Job Interview'' ($n =$ 15, 17\%), ``Meet a new roommate'' ($n =$ 14, 16\%), and then followed by ``Confrontation'' ($n =$ 12, 14\%), ``Asking Someone Out'' ($n =$ 11, 12\%), ``Public Speaking'' ($n =$ 10, 11\%). Less commonly selected scenarios included ``Auditioning for a Performance'' ($n =$ 6, 7\%) and ``Mentoring Students as a Teaching Assistant'' ($n =$ 2, 2\%). The observed usage distribution suggests that participants gravitated toward everyday social contexts tied to common anxiety triggers, consistent with the literature on social anxiety~\cite{hur2020social, leary1997social}.

The mean number of conversational exchanges during script development was 12.64 ($SD =$ 6.15), reflecting diverse engagement patterns: some participants preferred short, iterative exchanges, while others provided detailed descriptions in fewer turns. During the IE exercises, the average response length was 73 words ($SD =$ 81), ranging from 0 to 671 words, suggesting that participants' engagement levels and comfort with providing detailed input varied significantly. 

\subsection{Bridging Real-Life Connections and Evoking Anticipatory Anxiety}
\label{sec:strength}

Participants highlighted how \tool helped them \textbf{prepare for real-life anxiety-inducing situations} in ways that mirrored established IE practices. For example, P7 explained, \textit{``It helped me prepare for meeting my new roommate to avoid catastrophizing the scenario because I had known that I was likely to start doing that.''} Similarly, P14 noted, \textit{``It helped me a lot with an interview because it was planned and I knew what to do beforehand because of the tool.''} These comments suggest that personalized, LLM-generated scripts can reinforce the real-life applicability of IE by providing concrete practice for upcoming events, echoing prior findings that structured, forward-looking exercises can reduce avoidance behaviors~\cite{davidson2004fluoxetine,ledley2009comprehensive}. 

Similar to participants' comments, MHPs also emphasized \tool's \textbf{capacity to create vivid and realistic narratives}. For example, MHP4 commended the platform for its ability to create scenarios that felt real, noting that it\textit{``brought me right into that scenario,''} that the scenario \textit{``was a very descriptive''} and \textit{``really caught my level of anxiety and made it real.''} This realism was further enriched by the incorporation of sensory details. By engaging multiple senses---sight, hearing, smell, taste, touch, and emotional cues---the tool appeared to strengthen users’ emotional connection to the scenario. As MHP3 explained, \textit{``\tool incorporated the five senses as well as the emotional response to help people to bring attention to how our anxiety is often fostered by all of those different factors, not just the thoughts and emotions.''} Indeed, it is important to support immersiveness for IE practices, as previous research highlighted, as such vividness helps users relate more deeply to the content~\cite{heimberg2006managing}.

Moreover, both participants and MHPs recognized that a strength of \tool lies in \textbf{the deliberate elevation of anxiety to a ``window of tolerance.''} These concepts are grounded in psychotherapy literature, which emphasizes the need to raise discomfort to a therapeutic threshold---enough to foster engagement but not so high as to overwhelm~\cite{heimberg2006managing}. MHP1 noted that \textit{``initially, after listening to the recording,''} while \tool's text-to-audio feature might cause \textit{``a small spike in their anxiety,''} this elevation \textit{``stay within like a beneficial window''} for therapeutic progress.
The alignment of IE sessions with this ``window of tolerance'' reflects one feasibility of an LLM-enabled tool: \textbf{dynamically adjusting scenario intensity based on user input.} This feature can be especially helpful for individuals who struggle with traditional IE ``homework,'' since they receive real-time guidance to manage anxiety. By combining immersive scenario creation, multi-sensory prompts, and carefully calibrated stress levels, \tool is promising to reduce the burden on the user’s imagination and motivation, potentially enhancing compliance. 

Furthermore, MHPs emphasized that a controlled increase in anxiety is essential for therapeutic progress, with MHP1 noting that such \textit{``imaginal exercise is really good for those kinds of future-oriented fears and thoughts.''} This suggests that by keeping exposure challenging yet manageable, \tool aligns with the core principles of Imaginal Exposure (IE) within a safe, structured environment.

\subsection{The Need for Adaptivity and Continuity in LLM-enabled Imaginal Exposure Tools}
\label{sec:weakness}

While our evaluation shows promising feasibility, participants and MHPs identified some limitations that should be addressed for future effective use. One concern raised by MHPs involves \tool's \textbf{inability to build upon users' previous inputs}. While \tool briefly screens for past events by asking, \textit{``have you experienced anything like this before,''} MHP1 argued that the tool fails to truly \textit{``dig deeper''} into the history that contributes to \textit{``a higher level of anxiety than you think is warranted.''} This lack of longitudinal context creates a reset-based interaction; without carrying this deeper history forward, the dialogue remains at a surface level, preventing users from uncovering the underlying fears driving their current distress. MHP2 further highlighted the lack of customization, \textit{``I was less conscious about the facial expressions of people, but I was more conscious of other stimuli like auditory.''} This comment illustrates that while \tool achieves a baseline of personalization, there is an opportunity to further refine these individual triggers, and \tool could benefit from more granular control to better capture the nuanced complexity of users’ social anxiety. P16 suggested that relatability could be enhanced by offering \textit{``a more extensive selection or the ability to create personalized scenarios to improve relatability and effectiveness.''} Similarly, P6 criticized the lack of dynamism, commenting, \textit{``I felt like it just regurgitated what I said to it. It made me think through what a situation would look like but did not teach me about the implications of that situation.''} 
These findings echo concerns that a purely algorithmic response can feel mechanical, limiting the emotional connection essential for therapeutic engagement~\cite{jensen2020congruence}.

\section{Discussion \& Conclusion}
\textbf{Discussion.} Our findings reveal the potential of LLM-enabled tools to fill significant gaps in traditional IE practice for SA. 
\tool demonstrated the ability to generate vivid, multi-sensory scripts that support real-life preparation, maintain anxiety within a therapeutic range, and adapt to diverse triggers—capabilities that address several known barriers to IE homework adherence~\cite{sharma2023cognitivereframingnegativethoughts}. These results suggest that LLM-enabled tools can serve as a valuable bridge between therapy sessions, providing the structured support that many individuals need to engage effectively with IE exercises. However, the limitations our study identified---particularly around continuity and personalization---raise important questions about the appropriate role of LLMs in therapeutic contexts. Should LLM-enabled tools function as passive script generators, or evolve into more adaptive systems capable of tracking therapeutic progress over time? If the latter, new mechanisms are needed to ensure responsible and effective operation, especially given concerns about bias, privacy, and the quality of therapeutic guidance~\cite{Wang2026}. 
Our findings suggest that a hybrid model may be most appropriate, in which therapists remain meaningfully involved---for example, by reviewing LLM-generated scripts or providing context-specific guidance when needed. Future work should investigate how such collaborative workflows can be practically structured. This model would allow LLMs to complement, rather than replace, therapist expertise, maintaining human oversight while leveraging automation to reduce the burden of IE homework. The tension between structured guidance and user autonomy is another important design consideration. Effective IE homework requires active engagement, yet rigid scripts can reduce the sense of ownership and limit deeper introspection~\cite{Bhattacharjee2022}. Conversely, offering too many pathways may overwhelm users or dilute the therapeutic focus. Future work should investigate adaptive prompting strategies that balance structure with flexibility, potentially informed by real-time indicators of user engagement and anxiety levels. 

Together, these findings yield three design implications for future LLM-enabled IE tools: a therapist-in-the-loop model to maintain clinical oversight, longitudinal continuity to enable deeper therapeutic engagement, and adaptive prompting strategies to balance structure with user autonomy.

\textbf{Limitations.} 
Our work has several limitations. First, the design relies on prompt engineering techniques, and further investigation into more sophisticated approaches (e.g., fine-tuning or multi-modal integrations) is needed to refine the tool’s adaptability and depth. Furthermore, our participant pool was relatively small, limiting the generalizability of our findings. More human-subject evaluations are needed to examine the potential and limitations of LLM-enabled tools in various contexts and for different (sub)populations.

\textbf{Conclusion. }
In this study, we co-designed \tool with domain experts (mental health professionals) to explore how LLM-enabled tools could facilitate imaginal exposure exercises for social anxiety. Through a formative study with five mental health professionals and a user study with 19 participants, we found that LLM-generated IE scripts can help users prepare for real-life anxiety-provoking situations, produce vivid multi-sensory narratives aligned with clinical IE practices, and maintain anxiety within a therapeutically beneficial range. We also identified challenges in contextual continuity and personalization that should be addressed for LLM-enabled tools to achieve deeper therapeutic value. 

\begin{acks}
We thank our participants for their time and valuable feedback. We also thank the reviewers for their constructive reviews.
\end{acks}


\bibliographystyle{ACM-Reference-Format}
\bibliography{ref}

\appendix 

\newpage
\section{Appendices}
\label{sec:appendices}

\subsection{Survey Details}
\label{sec:survey_details}
\textbf{Social Anxiety}

1) Social Phobia Inventory (SPIN)~\cite{connor2000psychometric} is a validated scale used to assess the severity of social anxiety symptoms. The SPIN consists of 17 items, each rated on a scale from 0 (not at all) to 4 (extremely), resulting in total scores ranging from 0 to 68. Scoring involves summing all the items, with a score above 20 suggesting the possibility of social anxiety. In research contexts, this threshold has been effective in distinguishing between individuals with social phobia and those in a control group. Items include a series of statements that respondents rate to reflect their experiences and feelings in social situations, including: 
\begin{enumerate}
    \item I am afraid of people in authority.
    \item I am bothered by blushing in front of people.
    \item Parties and social events scare me.
    \item I avoid talking to people I don’t know.
    \item Being criticized scares me a lot.
    \item Fear of embarrassment causes me to avoid doing things or speaking to people.
    \item Sweating in front of others causes me distress.
    \item I avoid going to parties.
    \item I avoid activities in which I am the center of attention.
    \item Talking to strangers scares me.
    \item I avoid having to give speeches.
    \item I would do anything to avoid being criticized.
    \item Heart palpitations bother me when I am around people.
    \item I am afraid of doing things when people might be watching.
    \item Being embarrassed or looking stupid are among my worst fears.
    \item I avoid speaking to anyone in authority.
    \item Trembling or shaking in front of others is distressing to me.
\end{enumerate}

\begin{table}[h]
\centering
\caption{SPIN Scoring and Symptom Severity}
\begin{tabular}{cl}
    \toprule
    \textbf{Score} & \textbf{Symptom Severity} \\
    \midrule
    0 - 20 & None \\
    21 - 30 & Mild \\
    31 - 40 & Moderate \\
    41 - 50 & Severe \\
    51 - 68 & Very severe \\
    \bottomrule
\end{tabular}
\end{table}

2) Patient Health Questionnaire-9 (PHQ-9)~\cite{kroenke2001phq} is a validated scale used to assess the severity of depression symptoms. The PHQ-9 consists of 9 items, each rated on a scale from 0 (not at all) to 3 (nearly every day), resulting in total scores ranging from 0 to 27. Scoring involves summing all the items, with thresholds categorizing depression severity. Items include a series of statements reflecting the frequency of depressive symptoms over the past two weeks:
\begin{enumerate}
    \item Little interest or pleasure in doing things.
    \item Feeling down, depressed, or hopeless.
    \item Trouble falling or staying asleep, or sleeping too much.
    \item Feeling tired or having little energy.
    \item Poor appetite or overeating.
    \item Feeling bad about yourself—or that you are a failure or have let yourself or your family down.
    \item Trouble concentrating on things, such as reading the newspaper or watching television.
    \item Moving or speaking so slowly that other people could have noticed? Or the opposite—being so fidgety or restless that you have been moving around a lot more than usual.
    \item Thoughts that you would be better off dead or of hurting yourself in some way.
\end{enumerate}

\begin{table}[h]
\centering
\caption{PHQ-9 Scoring and Symptom Severity}
\begin{tabular}{cl}
    \toprule
    \textbf{Score} & \textbf{Symptom Severity} \\
    \midrule
    0 - 4 & Minimal \\
    5 - 9 & Mild \\
    10 - 14 & Moderate \\
    15 - 19 & Moderately severe \\
    20 - 27 & Severe \\
    \bottomrule
\end{tabular}
\end{table}

3) Generalized Anxiety Disorder-7 (GAD-7)~\cite{spitzer2006brief} is a validated scale used to assess the severity of generalized anxiety symptoms. The GAD-7 consists of 7 items, each rated on a scale from 0 (not at all) to 3 (nearly every day), resulting in total scores ranging from 0 to 21. Scoring involves summing all the items, with thresholds categorizing anxiety severity. Items include a series of statements reflecting the frequency of anxiety symptoms over the past two weeks:
\begin{enumerate}
    \item Feeling nervous, anxious, or on edge.
    \item Not being able to stop or control worrying.
    \item Worrying too much about different things.
    \item Trouble relaxing.
    \item Being so restless that it is hard to sit still.
    \item Becoming easily annoyed or irritable.
    \item Feeling afraid as if something awful might happen.
\end{enumerate}

\begin{table}[h]
\centering
\caption{GAD-7 Scoring and Symptom Severity}
\begin{tabular}{cl}
    \toprule
    \textbf{Score} & \textbf{Symptom Severity} \\
    \midrule
    0 - 4 & Minimal \\
    5 - 9 & Mild \\
    10 - 14 & Moderate \\
    15 - 21 & Severe \\
    \bottomrule
\end{tabular}
\end{table}

\textbf{Demographic background}

1) \textit{Gender.} Participants were given the options of ``Woman,'' ``Man,'' ``Non-binary,'' and ``Prefer not to answer.''

2) \textit{Education.} Participants were asked about their highest level of education using the question, \textit{``What is the highest degree or level of school you have completed?''} The response options included: ``High school graduate,'' ``Associate degree,'' ``Bachelor’s degree,'' ``Master’s degree,'' ``Doctorate or professional degree,'' or ``Prefer not to answer.''

3) \textit{Race.} Participants first were asked if they considered themselves as Hispanic or Latino (and allowed to select option from the following list: ``Yes,'' ``No,'' and ``I prefer not to answer''), and then could select from the options for race: ``White,'' ``African American or Black,'' ``American Indian or Alaskan Native,'' ``Native Hawaiian or Pacific Islander,'' ``Asian,'' ``Other,'' and ``Prefer not to answer.''

\subsection{Interview Codebook}

\begin{table}[h]
\centering
\caption{Codebook for Bridging Real-Life Connections and Evoking Anticipatory Anxiety}
\footnotesize
\begin{tabularx}{\linewidth}{p{1.5cm}p{3cm}X}
    \toprule
    \textbf{Theme} & \textbf{Description} & \textbf{Example} \\
    \midrule
    Vivid Scenario Recreation & The tool recreates real-life scenarios, helping users prepare for anxiety-inducing situations. & P7: \textit{``It helped me prepare for meeting my new roommate...''}  
    MHP4: \textit{``It brought me right into that scenario... it was a very descriptive, really good job with that...''} \\
    \midrule
    Manageable Anxiety Elevation & Temporarily increases anxiety within a ``window of tolerance,'' fostering therapeutic engagement without overwhelming users. & MHP1: \textit{``Initially, after listening to the recording, there might be a small spike in their anxiety... but it is gonna stay within like a beneficial window.''} \\
    \midrule
    Need for Continuity & The chatbot interactions lack continuity, requiring users to re-enter information and limiting deeper exploration. & MHP4: \textit{``There is a history that we are not digging deeper...''} \\
    \midrule
    Lack of Adaptivity & The tool’s rigid and repetitive nature reduces engagement and limits effectiveness. & P6: \textit{``I felt like it just regurgitated what I said to it... did not teach me about the implications of that situation.''} \\
    \bottomrule
\end{tabularx}
\end{table}

\end{document}